\documentclass[review]{elsarticle}

\usepackage{amssymb}

\journal{Nuclear Inst. and Methods in Physics Research, A}

\usepackage{HGTD_TDR-defs}
\usepackage{multirow} 
\usepackage{rotating}
\usepackage{mathrsfs}
\usepackage{booktabs}
\usepackage{longtable}
\usepackage{placeins}
\usepackage{amssymb}
\usepackage{upgreek}
\usepackage{svg}
\usepackage{comment}
\usepackage{pdfpages}
\usepackage{siunitx}
\usepackage{lineno}
\usepackage{float}
\usepackage{lineno}
\usepackage{tikz}
\usepackage[percent]{overpic}
\usepackage{caption}
\usepackage{subcaption}
\usepackage{makecell}
\usepackage[colorinlistoftodos]{todonotes}
\title{\boldmath Performance of neutron and proton irradiated AC-LGAD sensors}
 
\author[1]{G.~Stage}
\author[1]{A.~Borjigin}
\author[1]{J.~Ding}
\author[1]{M.~Davis}
\author[1]{S.~Beringer}
\author[2]{M.~Gignac}
\author[1]{F.~McKinney-Martinez}
\author[1]{S.M.~Mazza$^*$}
\author[1]{A.~Molnar}
\author[3]{J.~Ott}
\author[1]{H.F.-W.~Sadrozinski}
\author[1]{B.~Schumm}
\author[1]{A.~Seiden}
\author[1]{T.~Shin}
\author[1]{M.~Wilder}
\author[4]{G.~Kramberger}
\author[4]{I.~Mandic}
\author[5]{S.~Seidel}
\author[5]{J.~Si}
\author[5]{R.~Novotny}

\affiliation[1]{SCIPP, University of California Santa Cruz, 1156 High Street, Santa Cruz (CA), US}
\affiliation[2]{SLAC, 2575 Sand Hill Rd, Menlo Park, CA 94025}
\affiliation[3]{Department of Electrical and Computer Engineering, University of Hawaii at Manoa, 2540 Dole Street, Honolulu HI-96822, USA}
\affiliation[4]{IJS, Jamova cesta 39, 1000 Ljubljana, Slovenia}
\affiliation[5]{Department of Physics and Astronomy, University of New Mexico, 210 Yale Blvd. NE, Albuquerque, NM 87106}

\usepackage{sfmath}

\begin{document}

\begin{frontmatter}

\begin{abstract}
Characterization of strip and pixel AC-LGAD devices with both laser TCT and probe station (IV/CV) will be shown on AC-LGADs irradiated with 1 MeV reactor neutrons at JSI/Ljubljana and with 400~MeV protons at FNAL ITA to fluences from 1e13~$n_{eq}/cm^2$ to a few times 1e15~$n_{eq}/cm^2$.
This study was conducted within the scope of the ePIC detector time of flight (TOF) layer R\&D program at the EIC, which will feature AC-LGADs with strip and pixel geometry. Sensors in the TOF layer will receive up to 1e13~$n_{eq}/cm^2$ fluence over the lifetime of the experiment.
\end{abstract}

\begin{keyword}
fast silicon sensors; irradiation; charge multiplication; AC-LGAD strips; charge sharing.
$^*$Corresponding author: simazza@ucsc.edu
\end{keyword}

\end{frontmatter}


\section{Introduction}
\label{sec:intro}
\begin{figure}[!hbt]
 \centering
 \includegraphics[width=0.9\columnwidth]{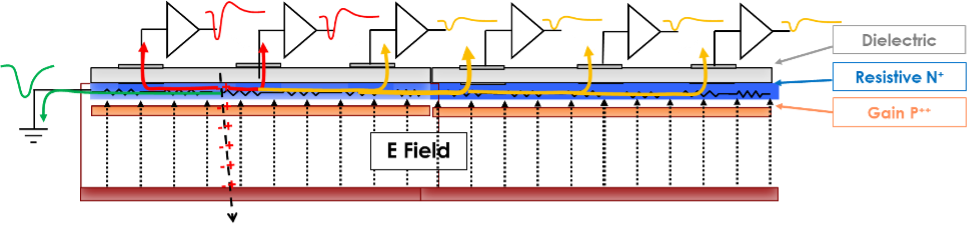}
 \caption{Cross section of an AC-LGAD showing common sensor layers and the signals shared by neighboring metal contacts. Red: direct, induced signal, yellow: pick-up from the N$\textsuperscript{+}$ layer and the signal traveling in the N$\textsuperscript{+}$ layer and collected in the ground contact (green). Schematic from~\cite{BISHOP2024169478}.}
 \label{fig:ACLGAD}
\end{figure}

Low Gain Avalanche Detectors (LGADs) have been established in the last 10 years as a fast semiconductor timing technology~\cite{Pellegrini:2014lki, Sadrozinski:2013nja} with a timing resolution of the order of tens of picoseconds. 
In LGADs' first large-scale experimental applications, the High Granularity Timing Detector (HGTD) in ATLAS~\cite{Mazza:2019dkn} and the MIP Timing Detector (MTD) in CMS~\cite{Ferrero:2022ynt}, the segmentation is limited to pads with about 1 mm pitch by consideration of power, fill-factor, and field uniformity. 
The fill-factor and uniformity are both solved for AC-LGAD technology, aka Resistive Silicon Detector (RSD)~\cite{BISHOP2024169478,Mandurrino}, which is based on a complete integration of four of the sensor layers in common sheets of the P-type bulk, the P$\textsuperscript{++}$ gain layer, the N$\textsuperscript{+}$ layer, and a dielectric sheet, separating the first three from the segmented metal readout contacts (Fig.~\ref{fig:ACLGAD}).
A signal originating in the bulk and amplified in the gain layer is then shared between several electronic channels, allowing signal location reconstruction with a resolution of a small ($<5\%$~\cite{MENZIO2024169526}) fraction of the readout pitch, also allowing an acceptable power density for high position and time resolution. 

AC-LGADs are the chosen sensor technology for the ePIC detector~\cite{AbdulKhalek:2021gbh} time of flight layer, which is composed of strips in the barrel region and pixels in the end-cap region. 
Over the lifetime of ePIC, a fluence up to 1e13~$n_{eq}/cm^2$ will be accumulated in the hottest regions of the sub-detector.
Strip and pixel AC$-$LGADs produced by Hamamatsu Photonics K.K. (HPK)~\cite{Hamamatsu} were tested before and after neutron and proton irradiation with 1 MeV reactor neutrons at JSI/Ljubljana and with 400~MeV protons at FNAL ITA to fluences on the order of 1e13~$n_{eq}/cm^2$ to a few times 1e15~$n_{eq}/cm^2$.
The sensors were characterized with a probe station for current versus voltage (IV) and capacitance over voltage (CV) to probe the breakdown and degradation of the gain layer with irradiation.
Then, a selection of sensors was tested with a focused laser TCT station to test the change of behavior after irradiation in terms of charge-sharing response, time of arrival, and rise time.

\section{Experimental}
\subsection{Sensors}

\begin{figure}[!hbt]
 \centering
 \includegraphics[width=0.7\columnwidth]{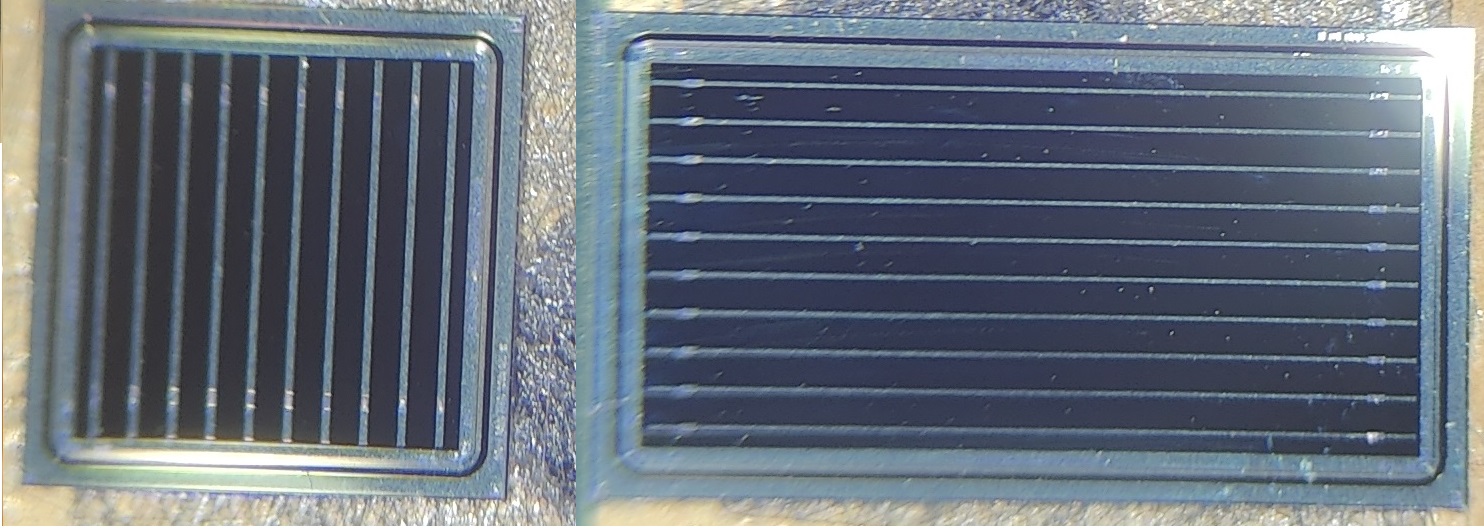}
 \caption{HPK AC-LGAD with 5~mm and 10~mm long strips on 500~µm pitch and 50~µm strip width.}
 \label{fig:ACLGAD_pic}
\end{figure}

The tested sensors were fabricated by HPK with funds from the eRD112 project to develop AC-LGAD detectors for the EIC~\cite{AbdulKhalek:2021gbh}. 
Fig.~\ref{fig:ACLGAD_pic} shows the layout of the approximately 5~mm wide AC-LGAD strip sensor with 5~mm and 10~mm long strips with 500 µm pitch and 50~µm strip width.

The wafers were fabricated with properties summarized in Tab.~\ref{tab:wafers}; each wafer included several pixel and strip sensors.
Two values for the N$\textsuperscript{+}$ sheet resistance (called “E-Type” and “C-Type” in the following) and two values for the dielectric thickness of the coupling capacitance were selected, resulting in four basic sensor combinations. 
An extensive laboratory and test beam characterization of non-irradiated sensors from this production was presented in \cite{BISHOP2024169478} and \cite{Dutta:2024ugh}, respectively.

\begin{table}[ht]
\centering
\begin{tabular}{|l|ccc|} 
\hline
\makecell{Wafer\\ \#} & \makecell{N$^+$ Resistance \\ {[}$\Omega$ / $\Box${]}} & 
\makecell{Dielectric C \\ (pF/$mm^2$)} & \makecell{Bulk Thickness \\ {[}$\mu$m{]}}\\ 
\hline
W02 & {E: 1600} & 240 & 50\\
\hline
W04 & C: 400 & 240 & 50 \\ 
\hline
W05 & {E: 1600}& 600 & 50\\ 
\hline
W08 & C: 400 & 600 & 50\\ 
\hline
W09& E: 1600 & 600 & 20 \\ 
\hline
\end{tabular}
\caption {Parameters of the tested HPK AC-LGAD wafers.}
\label{tab:wafers}
\end{table}

\subsection{IV and CV Measurements}
The sensors were tested in a probe station using needles connected to a HV power supply and an LCR meter. 
Measurements of current versus voltage (IV) were made to test the breakdown of the sensors, defined as the point of a dramatic increase in leakage current in the detector.
Since the probe station does not have cryogenic capabilities, this test is only reliable for non-irradiated or low-fluence sensors because of the high leakage current of highly irradiated samples.

The capacitance versus voltage (CV) was also measured for all the sensors to probe the degradation of the gain layer with radiation damage. The frequency of the LCR meter was set to 10~kHz for non-irradiated samples and to 1~kHz for the irradiated samples.
The depletion voltage of the gain layer ($V_{GL}$) can be calculated from the sharp variation in the $1/C^2$ over voltage distribution with the method explained in~\cite{Ferrero:2018fen}. 
$V_{GL}$ is then plotted as a function of fluence and fitted with $N_D=N_0e^{-c\phi}$ where $\phi$ is the fluence (units: $n_{eq}/cm^2$) and c (units: $cm^2/n_{eq}$) is the characteristic gain deactivation factor of the sensor.
The value of $V_{GL}$ is within 60~V, so it could be measured for all irradiated devices at room temperature.

\subsection{IR Laser TCT Measurements}
The charge collection measurements using TCT follow the method described in~\cite{BISHOP2024169478}. In short, the sensors are mounted on fast analog amplifier boards with 16 channels and a bandwidth of 1~GHz designed at Fermilab (FNAL)~\cite{Heller:2022aug} and read out by a fast oscilloscope (2 GHz, 20 Gs). An infrared (IR) 1064 nm pulsed laser with a pulse temporal width of 30 ps and a spot of 10-20~$\mu$m width mimics the response of a MIP in the silicon~\cite{Particular}. The IR laser cannot penetrate metal; therefore, the sensor behavior can be characterized only in between metal electrodes.
The response of the sensor as a function of laser illumination position is evaluated using X-Y moving stages. Waveforms are averaged for each position to decrease the effect of laser power fluctuations. 
The pulse maximum (Pmax)~\cite{Ott:2022itj}, rise time (10-90\%), and time of arrival (time of the pulse maximum, Tmax) are calculated for each position and plotted as a function of position. The Pmax distributions shown are normalized to the maximum value of each respective distribution for ease of comparison.

\subsection{Irradiation campaign}

\begin{figure}[!hbt]
 \centering
 \includegraphics[width=0.6\columnwidth]{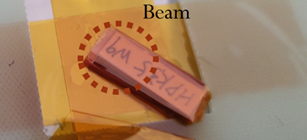}
 \caption{HPK AC-LGAD 2~cm long strip position with respect to the beam to perform an irradiation with a fluence gradient.}
 \label{fig:graded}
\end{figure}

The LGADs were irradiated without bias in the JSI research reactor of TRIGA type in Ljubljana, which has been used successfully in the past decades to support the development of radiation hard sensors~\cite{JSI_facility}.
The neutron energy spectrum and flux are well known.
The second set of LGADs was irradiated with 400~MeV protons at the FNAL ITA irradiation facility with a NIEL hardness factor of 0.65.
The fluence for both facilities is quoted in 1~MeV equivalent neutrons per cm$^2$ ($n_{eq}/cm^2$) by using Non-Ionizing Energy Loss (NIEL) scaling.
Fluences were between 1e12~$n_{eq}/cm^2$ and 1e15~$n_{eq}/cm^2$. 
Two of the large strip sensors were irradiated at FNAL ITA with a fluence gradient by positioning the beam off-center from the detector, as shown in Fig.~\ref{fig:graded}.
A series of test foils measured the fluence that varies from side to side by a factor of 5, from 4.4e14~$n_{eq}/cm^2$ to 7.8e13~$n_{eq}/cm^2$.

The uncertainty of the fluence at the TRIGA reactor is around $5\%$.
After irradiation, the devices were annealed for 80 min at 60~C. Afterward, the devices were kept in cold storage at -20~C to reduce further annealing.
Several 2x2 mm pixel AC-LGADs were made available for irradiation to high fluence. However, due to the limited availability of samples, a smaller number of strip sensors were irradiated only to modest fluences.

\section{Results and Discussion}

\subsection{Electrical characterization}
\begin{figure}[!hbt]
 \centering
 \includegraphics[width=0.48\columnwidth]{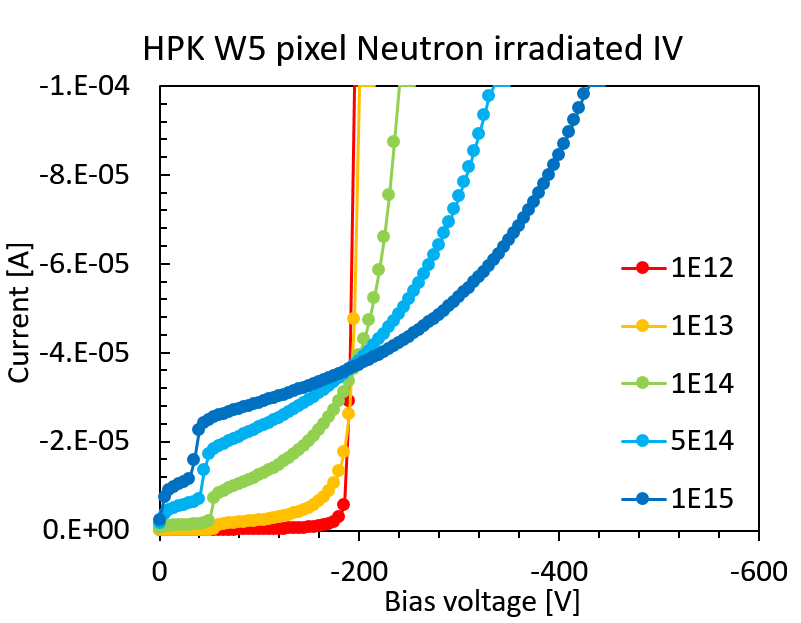}
  \includegraphics[width=0.48\columnwidth]{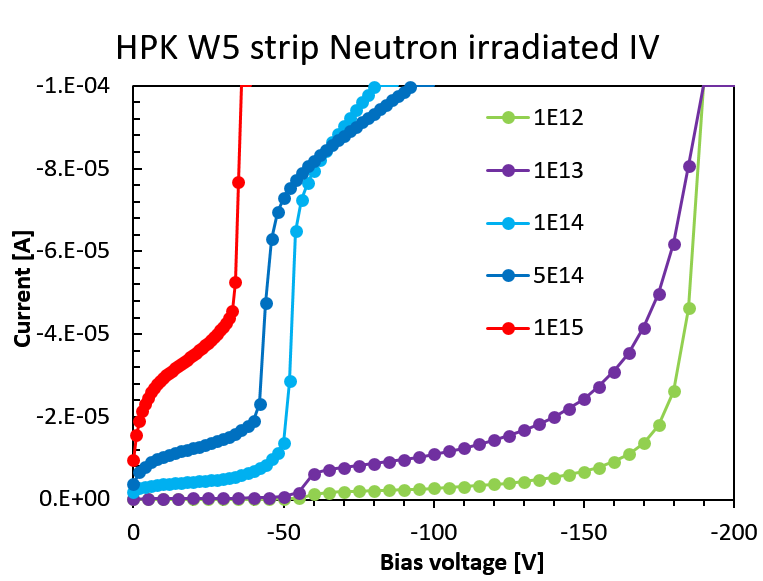}
 \caption{HPK AC-LGAD IV from W5 (50~$\mu m$ thickness) pixel (left) and strips (right) after neutron irradiation. The dark current increases as expected with fluence, and it's too high for larger strip sensors to be measured at room temperature.}
 \label{fig:pixel_IVs}
\end{figure}

IV and CV measurements of both strip and pixel AC-LGADs were executed at UC Santa Cruz after annealing. The IVs of a subset of the tested detectors are shown in Fig.~\ref{fig:pixel_IVs} before and after neutron irradiation. 
The sensors behave as expected, showing an increased dark current and higher breakdown voltage. Even though the fluence for the irradiated strips is lower, the current is significantly higher with respect to the pixels due to the increased area of the detectors.

\begin{figure}[!hbt]
 \centering
 \includegraphics[width=0.95\columnwidth]{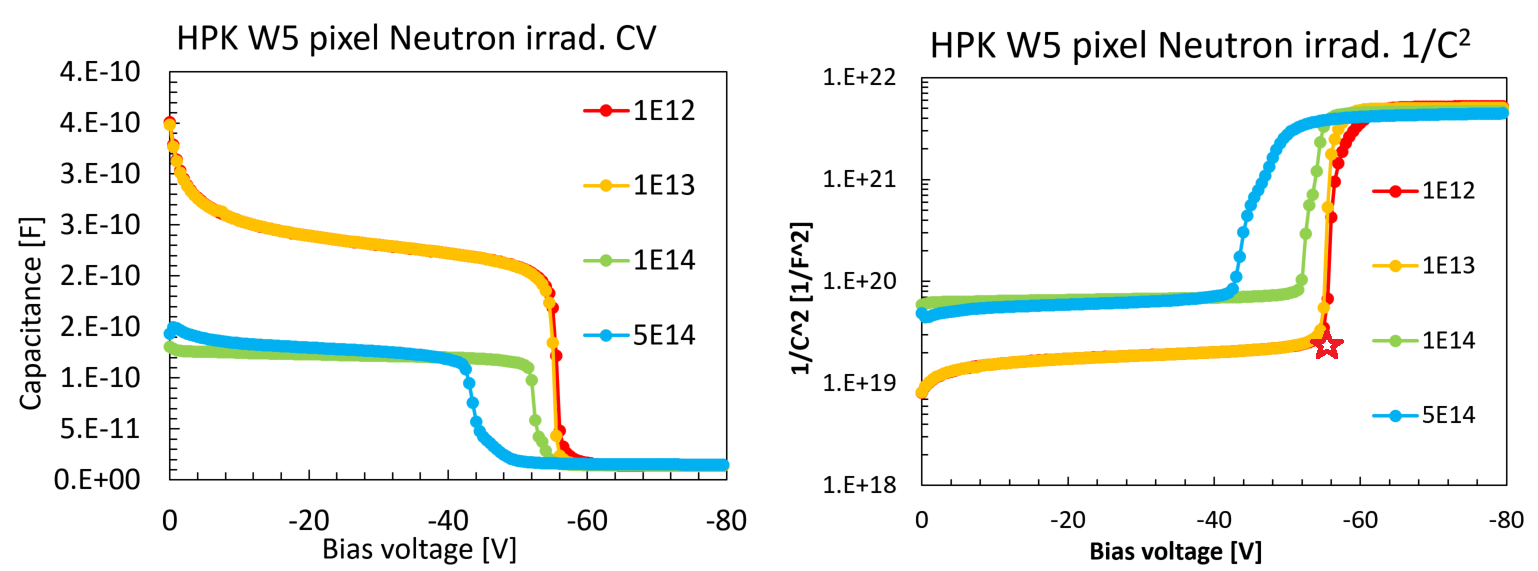}
 \caption{HPK AC-LGAD pixel CV (left) and 1/C$^2$ (right) from W5 (50~$\mu m$ thickness) after neutron irradiation. The gain layer depletion ($V_{GL}$, indicated for 1e12~$n_{eq}/cm^2$ by the red star in 1/C$^2$) is the sharp drop in capacitance after the gain layer is depleted; afterward, the bulk quickly depletes. As expected, $V_{GL}$ is lower in voltage for higher fluences.}
 \label{fig:pixel_CVs}
\end{figure}

\begin{figure}[!hbt]
 \centering
 \includegraphics[width=0.95\columnwidth]{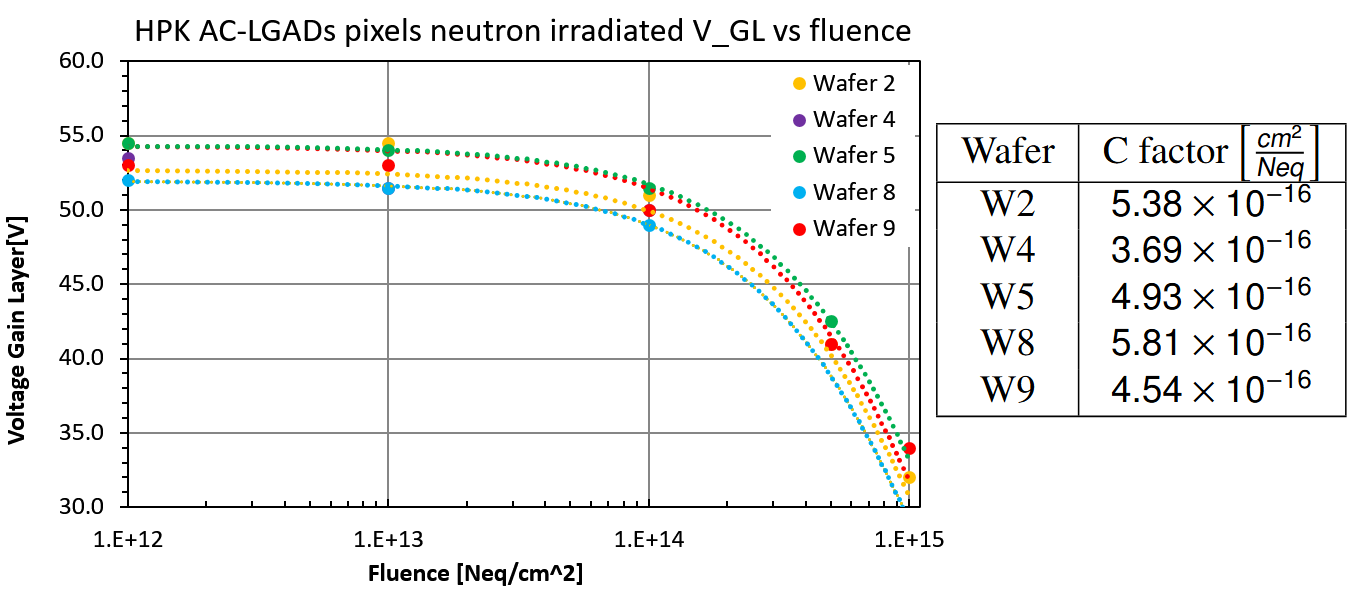}
 \caption{Distribution of $V_{GL}$ vs fluence for HPK AC-LGAD pixels of different wafers irradiated with 1~MeV neutrons. The function $N_D=N_0e^{-c\phi}$ is fitted to measure the c-factor for each wafer.}
 \label{fig:pixel_Vgl}
\end{figure}

The CV of pixel sensors from W2 and W9 after neutron irradiation is shown in Fig.~\ref{fig:pixel_CVs}. The variation of the gain layer depletion with irradiation from 55~V (1e12~$n_{eq}/cm^2$) to 35~V (highest fluence) is evident. From the inflection point $V_{GL}$, indicated as an example for 1e12~$n_{eq}/cm^2$ by the red star in Fig.~\ref{fig:pixel_CVs} (right), is measured and plotted against fluence in Fig.~\ref{fig:pixel_Vgl}.
The $V_{GL}$ vs fluence distribution from each wafer is fit with $N_D=N_0e^{-c\phi}$ to measure the c-factor for each wafer. 
The values are shown in the table in Fig.~\ref{fig:pixel_Vgl} (right).
Some wafers have a limited number of points, but generally, the c values vary between 4.5 and 5.5.
Similar c values were measured with the standard HPK ``HPK~3.1'' DC-LGAD~\cite{Padilla_2020}.
No dramatic difference in the behavior after neutron irradiation between DC-LGADs and AC-LGADs was observed.

\begin{figure}[!hbt]
 \centering
 \includegraphics[width=0.95\columnwidth]{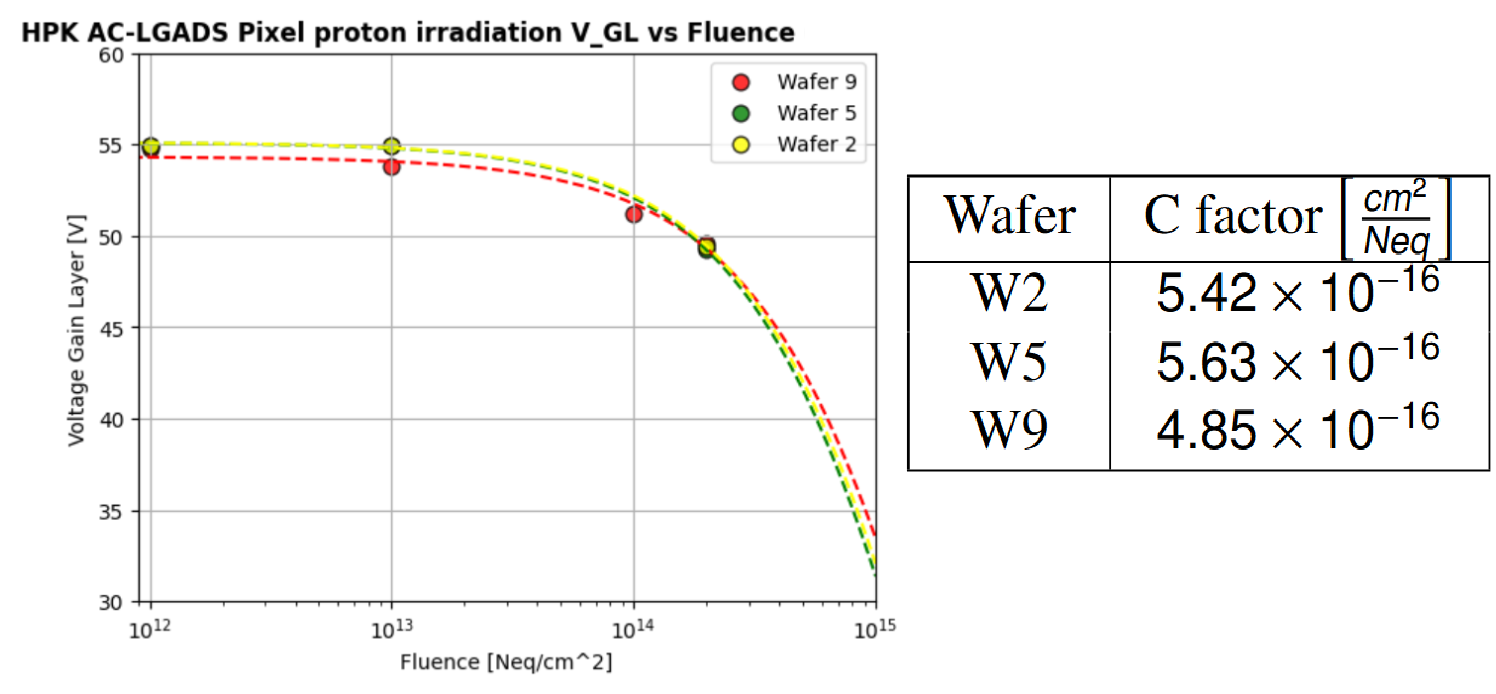}
 \caption{Distribution of $V_{GL}$ vs fluence for HPK AC-LGAD strips of different wafers irradiated with 400~MeV protons. The function $N_D=N_0e^{-c\phi}$ is fitted to measure the c-factor for each wafer.}
 \label{fig:proton_Vgl}
\end{figure}

The same analysis was performed for proton-irradiated sensors; the results are shown in Fig.~\ref{fig:proton_Vgl}. Unfortunately, fewer pixel and strip samples from the same wafers were available for proton irradiation. Therefore, the fluence was limited to 2e14~$n_{eq}/cm^2$, yielding sub-optimal fit quality to measure the c-factor. 
The c-factor from proton and neutron irradiation is similar, with a marginally higher value for proton irradiation. Previous studies showed how proton irradiation is more damaging for the gain layer degradation in LGADs at the same NIEL fluence~\cite{Mazza:2018jiz}.

\FloatBarrier
\subsection{Laser TCT characterization}

\begin{figure}[!hbt] 
 \centering
 \includegraphics[width=0.46\columnwidth]{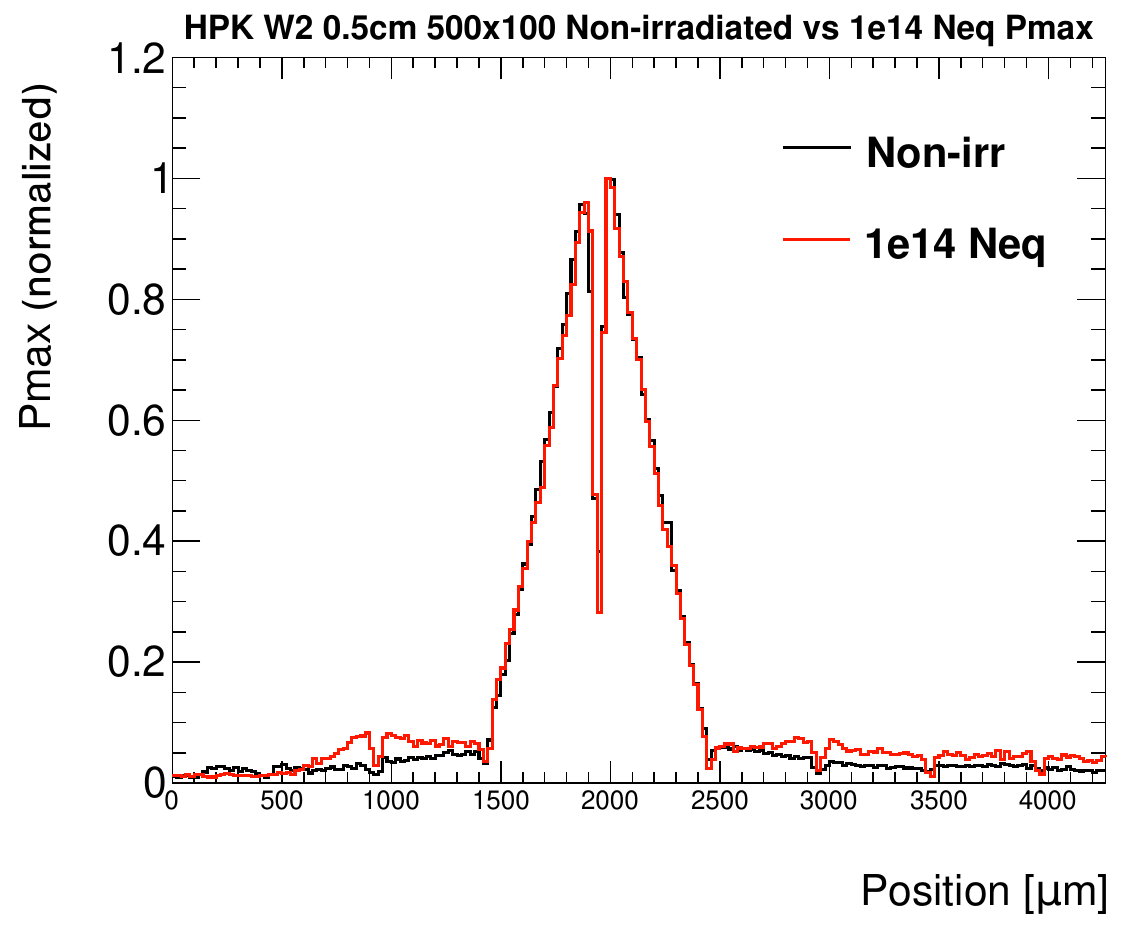}
 \includegraphics[width=0.49\columnwidth]{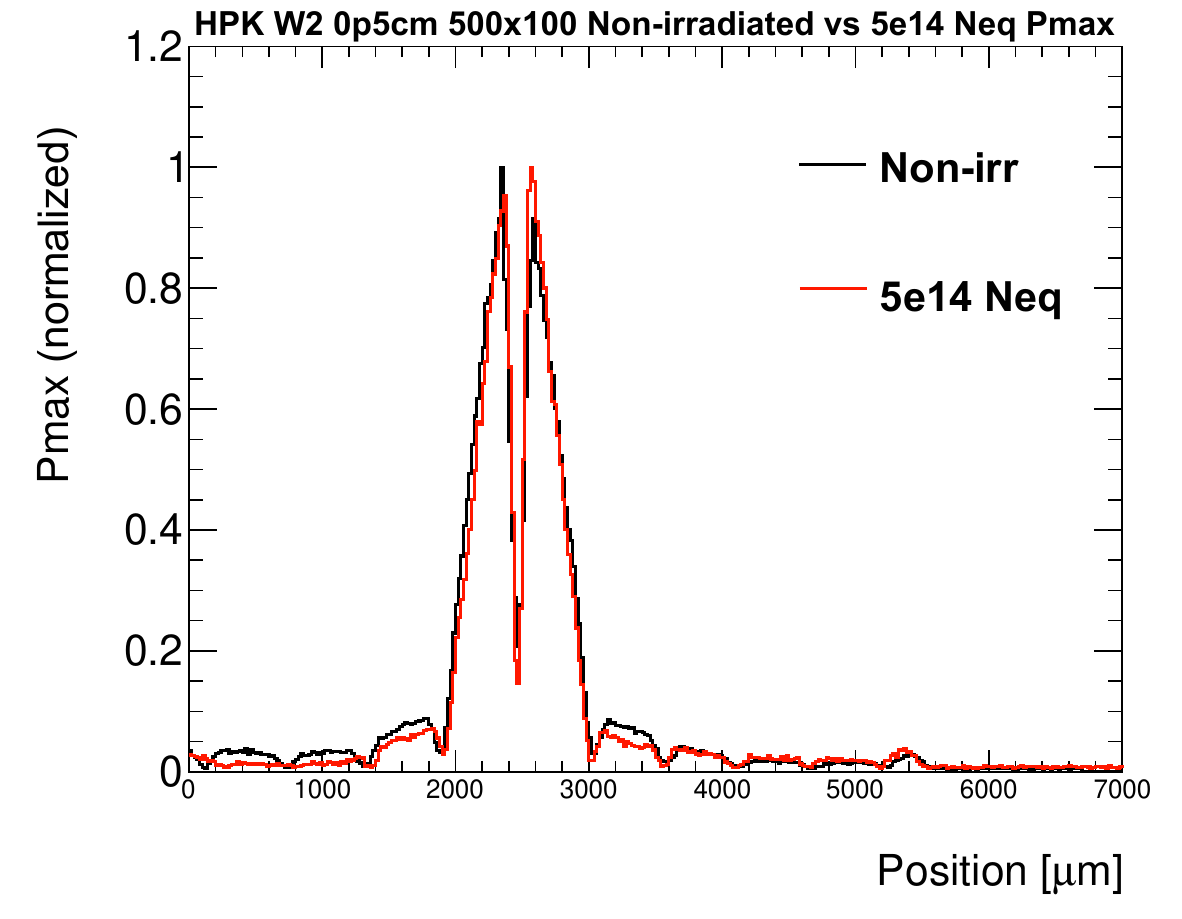}
 \caption{Pmax profile for an HPK AC-LGAD strip sensor from W2, 500~$\mu m$ pitch, 50~$\mu m$ (left) 100~$\mu m$ (right) strip width before and after neutron irradiation at 1e14~$n_{eq}/cm^2$ and 5e14~$n_{eq}/cm^2$. No change is seen in the charge-sharing distribution properties of the N$^+$ at first order.}
 \label{fig:pmax_profile}
\end{figure}

A selection of sensors were tested with the laser TCT system before and after neutron and proton irradiation.
The AC-LGAD strip sensor from W2 (50 um thickness, 500 um pitch, and 50-100~um strip width) will be presented for neutron irradiation. This type of sensor was tested before irradiation and after 1e14~$n_{eq}/cm^2$ and 5e14~$n_{eq}/cm^2$. 
The resulting Pmax distribution that characterizes the charge sharing is shown in Fig.~\ref{fig:pmax_profile} for 1e14~$n_{eq}/cm^2$ (left) and 5e14~$n_{eq}/cm^2$ (right); no change is observed in the charge sharing profile until the first neighboring strip. 
The maximum signal reached by the sensor is similar since the bias voltage is increased for the irradiated sensors. 
However, minor secondary effects in the distribution are observed after the first neighboring strip in both cases.

\begin{figure}[!hbt]
 \centering
 \includegraphics[width=0.46\columnwidth]{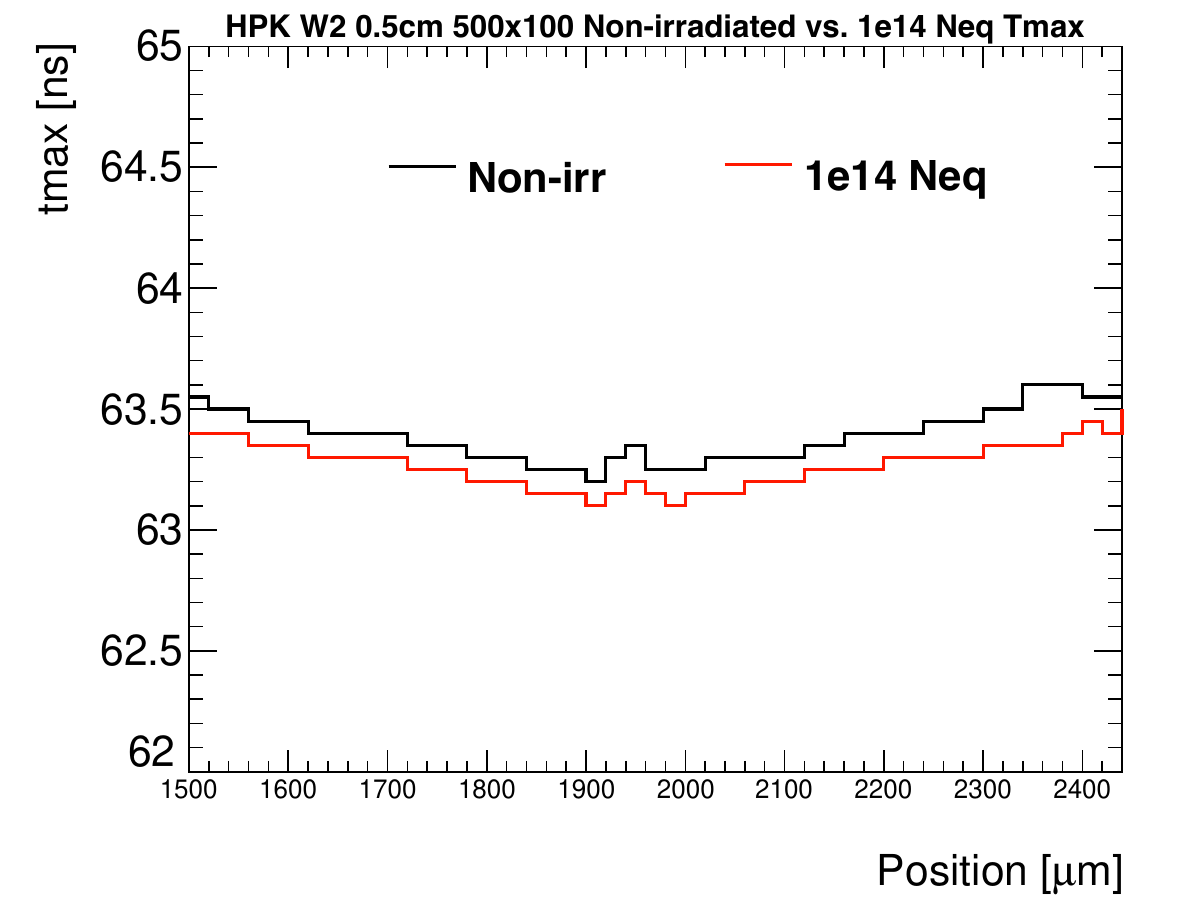}
 \includegraphics[width=0.46\columnwidth]{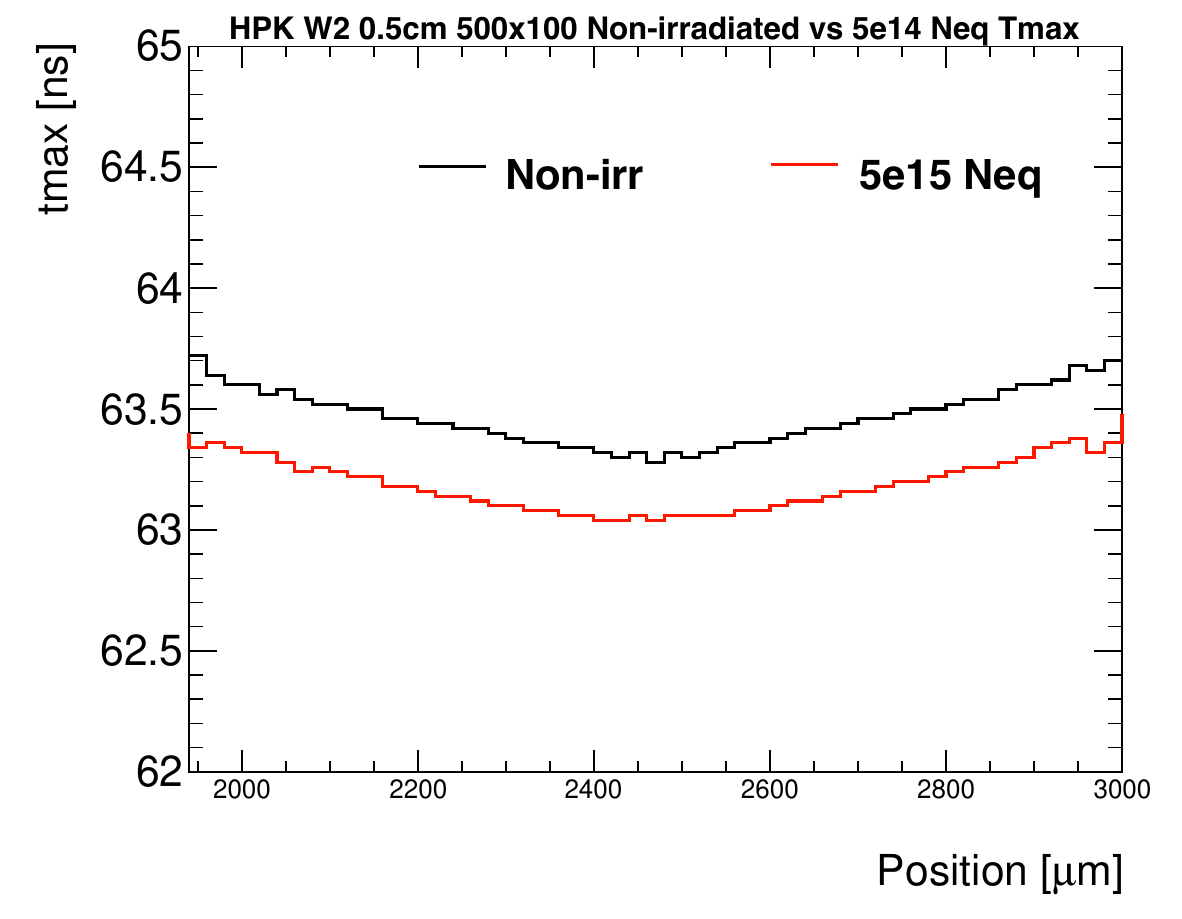}
 \caption{Tmax (time of arrival of the pulse) profile before and after neutron irradiation at 1e14~$n_{eq}/cm^2$ (left) and 5e14~$n_{eq}/cm^2$ (right). Only the time of arrival variation is relevant as the time delay is arbitrarily set on the oscilloscope.}
 \label{fig:time_profile}
\end{figure}

\begin{figure}[!hbt]
 \centering
 \includegraphics[width=0.46\columnwidth]{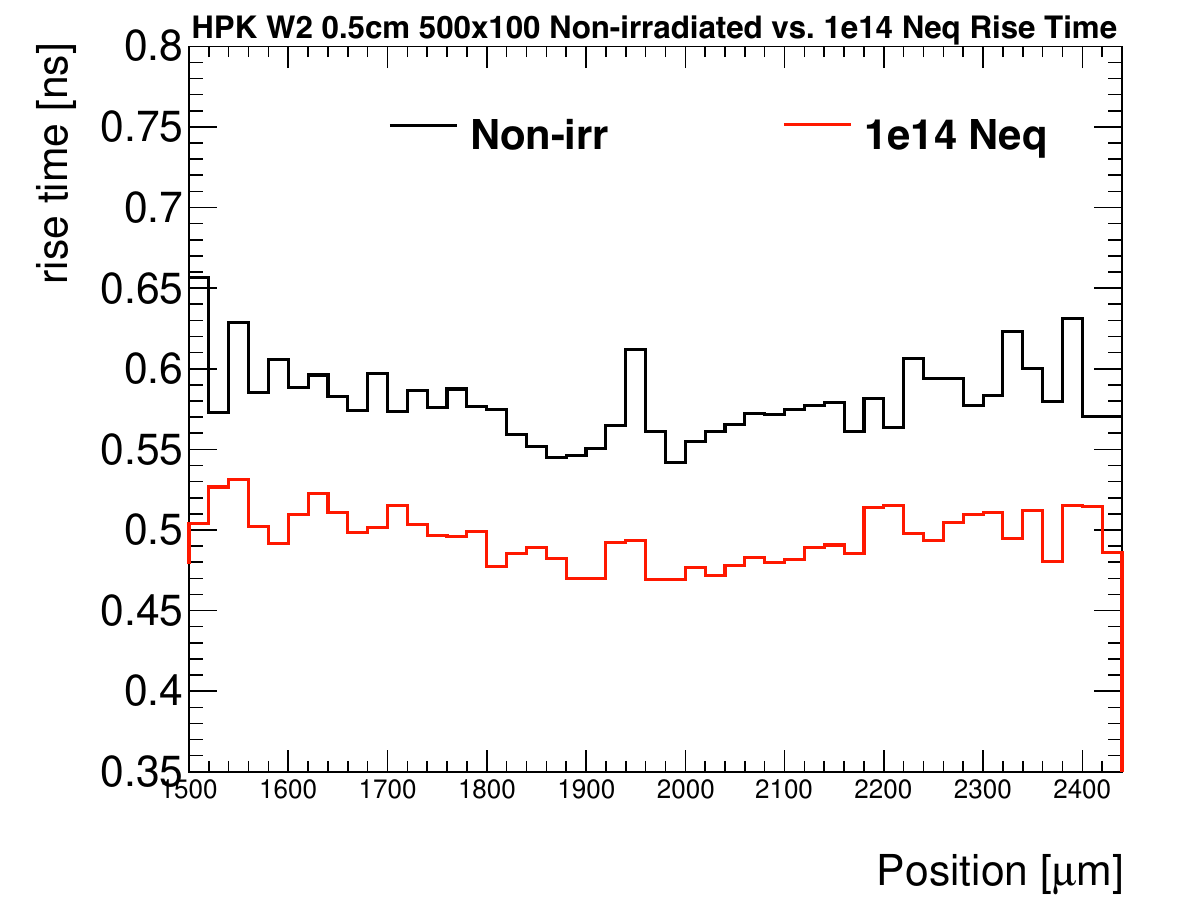}
 \includegraphics[width=0.46\columnwidth]{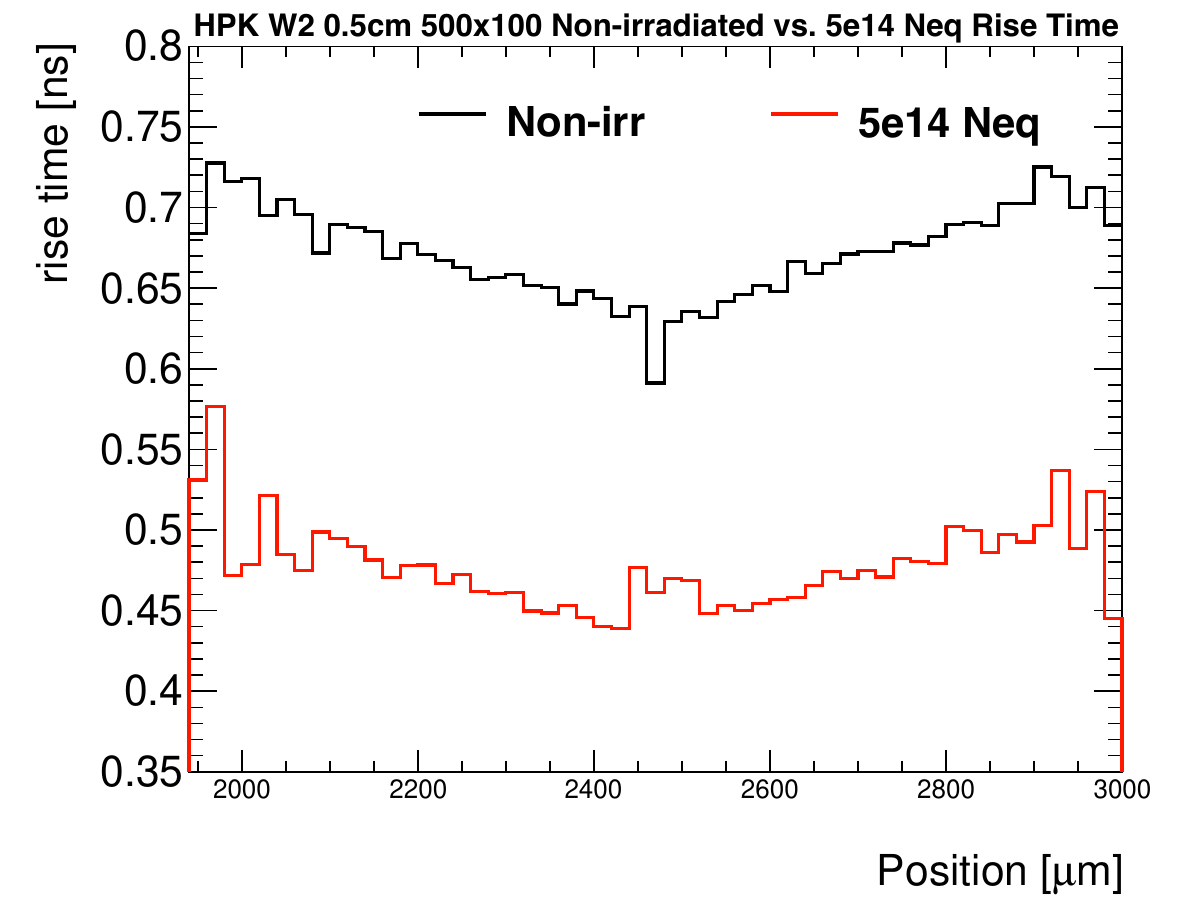}
 \caption{Rise time comparison between unirradiated and irradiated with neutrons at 1e14~$n_{eq}/cm^2$ (left) and 5e14~$n_{eq}/cm^2$ (right). The irradiation affects the rise time, and it's lower in both cases after irradiation. Before irradiation, the rise time for the 100~$\mu m$ wide strip is slightly larger, probably due to the increased input capacitance.} 
 \label{fig:rise_time_profile}
\end{figure}

Fig.~\ref{fig:time_profile} shows the time of arrival, and Fig.~\ref{fig:rise_time_profile} shows the rise time of the sensors before and after irradiation. 
The time of arrival follows the same distribution, showing no change in the signal propagation after irradiation. 
However, the rise time and the pulse in general are faster after irradiation. It's unclear what is causing this reduction and the cross-talk increase; a possible explanation is the increased conductivity of the surface due to irradiation causing a faster later charge evacuation through the N$^+$ layer.

\begin{figure}[!hbt]
 \centering
 \includegraphics[width=0.46\columnwidth]{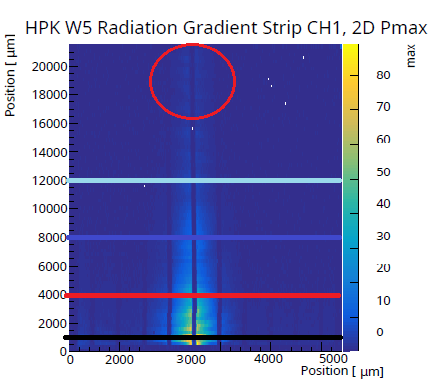}
  \includegraphics[width=0.48\columnwidth]{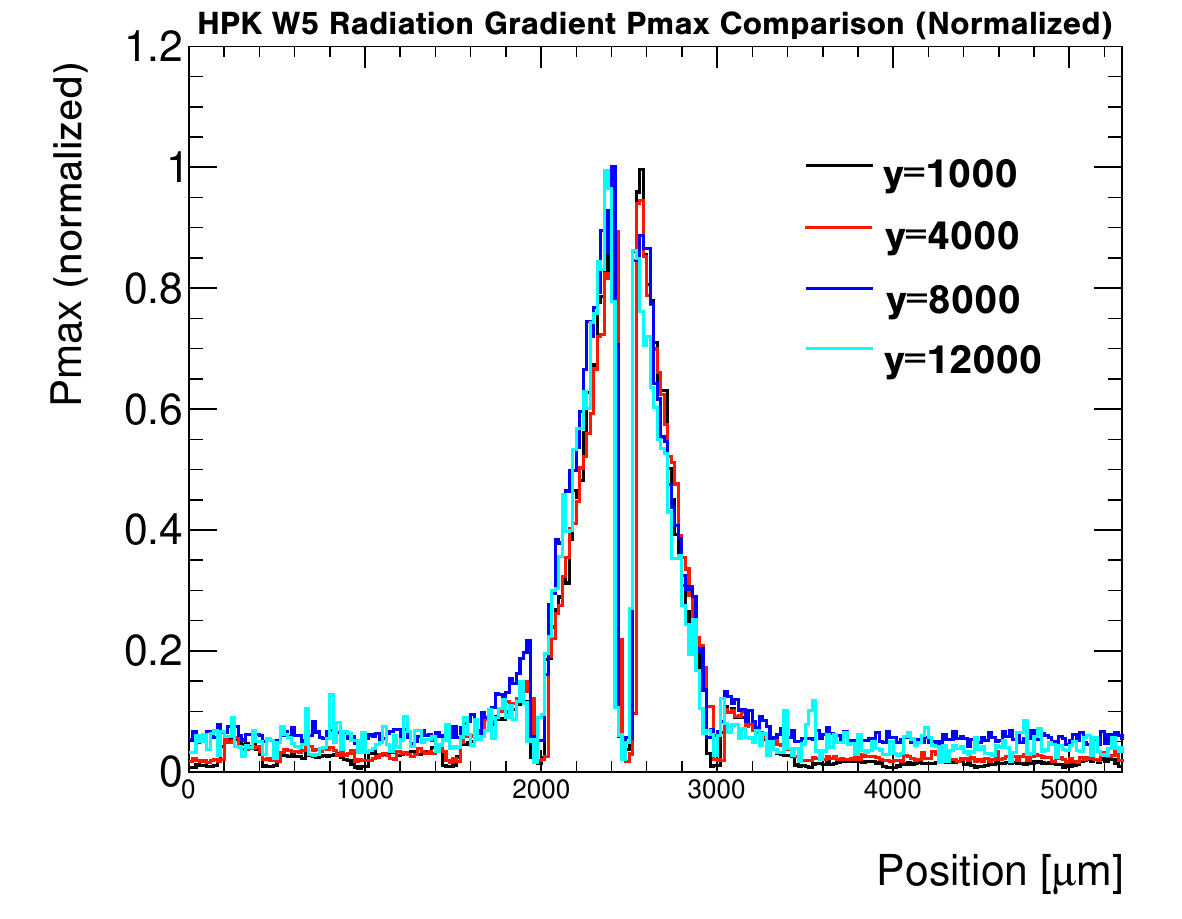}
 \caption{HPK AC-LGAD sensor from W5 2~cm length, 500~$\mu m$ pitch, 50~$\mu m$ strip width irradiated with protons with gradient fluence. (Right) HPK AC-LGAD strip full 2D TCT scan. (Left) normalized comparison of Pmax at different positions along the length of the strip, indicated by the colored lines in (right).}
 \label{fig:graded1}
\end{figure}

A full 2D scan of the sensor irradiated at FNAL with a fluence gradient (Fig.~\ref{fig:graded}) is shown in Fig.~\ref{fig:graded1} (left). The sensor works properly and shows a gain gradient proportional to the fluence; the beam was centered at the top (in the red circle) of Fig.~\ref{fig:graded1} (left).
A direct comparison of the Pmax profile perpendicular to the strip at different positions along the strip is shown in Fig.~\ref{fig:graded1} (right). No change is observed in the charge-sharing profile between the strip and the first neighbor across the sensor. 

\section{Conclusions}
The effect of neutron and proton radiation damage on HPK AC-LGADs was studied up to a fluence of a few times 1e15~$n_{eq}/cm^2$.
The gain layer degradation was measured with electrical characterization, and the results are in line with HPK's previous standard LGAD production.
The effect of irradiation on the charge-sharing mechanism was tested with the laser TCT, and no first-order change was observed in the Pmax profile and time of arrival. 
A secondary effect was observed on the long-distance charge sharing and pulse rise time.
This effect currently doesn't have a clear explanation; it could be an effect of N$^+$ changing resistivity due to irradiation damage or increased conductivity in the oxide/surface due to trapped charge.
Additional sensors will be tested at UC Santa Cruz to better understand the phenomena.

\tiny
\section{Acknowledgments}
We thank the technicians and students at SCIPP for the support and the time spent in the lab.
We acknowledge the collaboration with the KEK group (K. Nakamura et al.), the FNAL group (A. Apresyan et al.), the BNL group (A. Tricoli et al.), and the UIC group (Z. Ye, now at LBNL, et al.). 
The sensors were produced with funds from the ePIC erd112 effort for AC-LGADs development for the ePIC detector TOF layer.
The irradiation at the TRIGA reactor was covered by the European Union’s Horizon Europe Research and Innovation program under Grant Agreement No 101057511 (EURO-LABS).
This work was supported by the United States Department of Energy grants DE-SC0010107 and DE-SC0020255.

\bibliography{bib/TechnicalProposal,bib/hpk_fbk_paper,bib/HGTD_TDR,bib/SHIN,bib/nizam,bib/others}

\end{document}